\documentclass{article}
\usepackage{ijcai22}

\usepackage{times}
\usepackage{soul}
\usepackage{url}
\usepackage[hidelinks]{hyperref}
\usepackage[utf8]{inputenc}
\usepackage[small]{caption}
\usepackage{graphicx}
\usepackage{amsmath}
\usepackage{amsthm}
\usepackage{booktabs}
\usepackage{algorithm}
\usepackage{algorithmic}
\usepackage{subfigure}
\usepackage{multirow}
\usepackage{mathrsfs}

\newenvironment{sequation}{\begin{equation}\small}{\end{equation}}

\title{Debiasing Backdoor Attack: A Benign Application of Backdoor Attack in Eliminating Data Bias}

\author{
Shangxi Wu
\and
Qiuyang He\and
Yi Zhang\And
Jitao Sang
\affiliations
Beijing Jiaotong University\\
\emails
wushangxi@bjtu.edu.cn
}

\begin{document}

\maketitle

\begin{abstract}
    Backdoor attack is a new AI security risk that has emerged in recent years. Drawing on the previous research of adversarial attack, we argue that the backdoor attack has the potential to tap into the model learning process and improve model performance. Based on Clean Accuracy Drop (CAD) in backdoor attack, we found that CAD came out of the effect of pseudo-deletion of data. We provided a preliminary explanation of this phenomenon from the perspective of model classification boundaries and observed that this pseudo-deletion had advantages over direct deletion in the data debiasing problem. Based on the above findings, we proposed Debiasing Backdoor Attack (DBA). It achieves SOTA in the debiasing task and has a broader application scenario than undersampling.
\end{abstract}

\section{Introduction}
Backdoor attack~\cite{DBLP:journals/corr/abs-2007-08745} and 
adversarial attack are both traditional security 
risks in artificial intelligence. While adversarial attack is a 
technique to make a model produce wrong output results by adding a 
small human-designed perturbation, backdoor attack can make a model produce specified wrong 
output results by adding 
an artificially designed trigger. Since the backdoor attack was proposed, 
scholars mainly focused on the 
attack and defense algorithms of the backdoor attack. 
Major attack algorithms include BadNet~\cite{DBLP:journals/corr/abs-1708-06733} based on 
trigger training and TrojanNet~\cite{DBLP:conf/ndss/LiuMALZW018} 
based on model implantation. Later, other 
algorithms were proposed to enhance the stealthiness of backdoor 
attack~\cite{DBLP:conf/eccv/LiuM0020}. Nowadays, a large number 
of defense and detection 
algorithms are proposed according to the weaknesses of backdoor 
attack~\cite{DBLP:conf/acsac/GaoXW0RN19,DBLP:conf/raid/0017DG18,DBLP:conf/ijcai/ChenFZK19}.

The development of backdoor attack is 
similar to that of adversarial attack. In view of the 
research on the adversarial attack, we believe that 
the backdoor attack is worth exploring for model understanding. For example, in the study of 
adversarial training, those trained 
models not only have significant defensive effects against 
adversarial attack, but also significantly improve the 
interpretability of models~\cite{DBLP:journals/corr/GoodfellowSS14,DBLP:conf/icml/ZhangZ19}. 
In addition, it has been shown that 
adversarial attack can be used in benign applications, such as 
face privacy protection~\cite{DBLP:journals/corr/abs-2107-11986,DBLP:conf/mm/ZhangSZHSH20}, 
resisting malicious algorithms~\cite{DBLP:journals/tmm/ZhangSXWZSHY21} 
and adversarial data augmentation~\cite{DBLP:conf/mm/ZhangS20}.
It has been shown that backdoor 
attack can also be applied in benign scenarios. 
However, it has only been used in the field of security such as model 
protection~\cite{DBLP:conf/uss/AdiBCPK18,DBLP:conf/uss/JiaCCP21} based on the designability of the trigger. 

We would like to explore the properties present in backdoor attack 
and apply them to benign scenarios. Current measures of the 
effectiveness of trigger-based backdoor attack include ASR (Attack 
Success Rate), CAD (Clean Accuracy Drop), and stealthiness. 
We found that CAD was caused by pseudo-deletion by adding the 
trigger in backdoor attack, so as to make the model less generalizable. 
We managed to make a preliminary explanation 
of the CAD phenomenon by model classification boundaries. 
We proposed a new benign application based on backdoor attack 
by existing properties. Main contributions in this paper are shown as follows.

\begin{itemize}
    \item We found that the attacked model produces a pseudo-deletion effect on samples with triggers, and provided a preliminary explanation with classification boundaries.
    \item We found that such pseudo-deleted samples can affect the model classification boundaries and make better use of the training data than direct deletion when applying to undersampling scenarios.
    \item We proposed Debiasing Backdoor Attack, which has achieved SOTA in task of debiasing on CelebA and UTK Face. We also validated its extension to multi-class classification and the avoiding of security risks.
\end{itemize}

\section{Data Analysis}
CAD plays an important role in the design and 
evaluation of backdoor attack. 
CAD refers to the difference in accuracy 
between the attacked model and the normal model on clean inputs. 
Since the main difference between the backdoor-attacked 
model and the normal model only exists in training 
samples labeled with triggers, we decided to start from 
those samples and examine how the deleting or editing of 
those samples affect model on the training process 
and final results.

\subsection{Similarities and Differences between Backdoor Attack and Data Deletion on CAD}
\label{sec21}
Deleting data from the training set decreases the generalizability 
of the model, which also causes the accuracy drop. 
We want to compare the difference between the accuracy drop 
caused by data deletion and adding triggers.

\begin{figure}[!t]
    \begin{center}
       \includegraphics[width=1.0\linewidth]{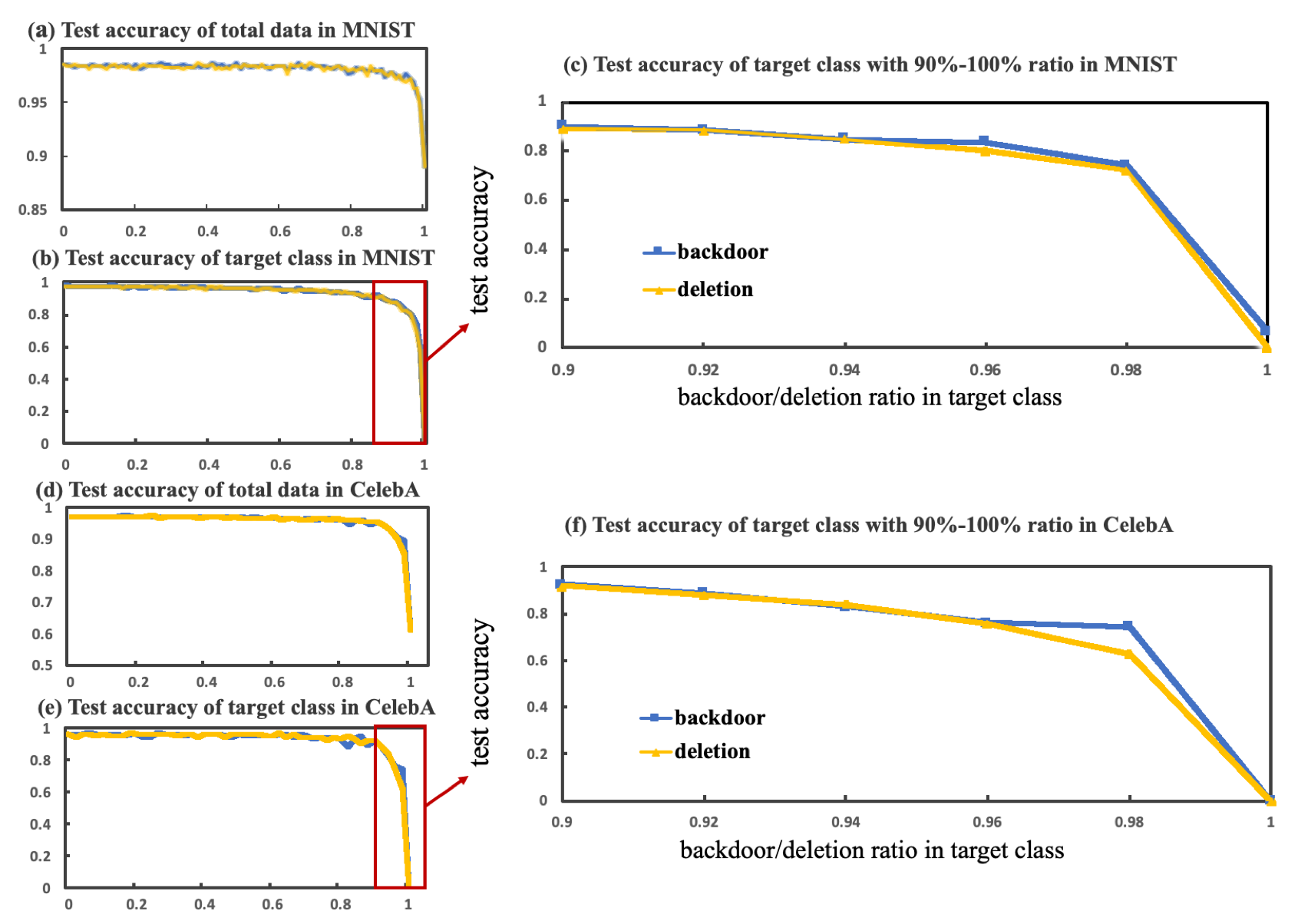}
    \end{center}
       \caption{Difference in test set accuracy between data 
       deletion and adding triggers. In MNIST, we deleted the 
       samples in class ``8". In CelebA we deleted the samples 
       with attribute ``Male".}
    \label{data21}
\end{figure}

Taking two tasks of classifying numbers in MNIST dataset and classifying gender 
in CelebA dataset as examples, we randomly deleted $p\%$ of data from a class or 
added $p\%$ of triggers to the class, and compared the accuracy of test set, as shown in Fig.~\ref{data21}.
It can be seen from Fig.~\ref{data21} that accuracy between 
the attacked model and the model with data deletion is close. 
Therefore, we speculated that adding a trigger brings a similar 
effect with data deletion. Nonetheless, at higher $p\%$, the accuracy 
degradation caused by the backdoor attack will be slightly lower 
than that of the data deletion. In particular, when 100\% of the 
training data of the target class is deleted, the accuracy of the 
trained model is 0\% for that class when tested. However, when 
adding triggers to 100\% of the data in that class, the model's 
accuracy is still about 5\% for that class.

Then we discussed how the similar effect occurred while adding a 
trigger in training process. By measuring the loss of samples 
with a trigger and normal samples separately, we verified whether 
there was a difference between samples with a trigger and normal 
samples of their contribution throughout the training process.

\begin{figure}[!t]
    \begin{center}
       \includegraphics[width=.9\linewidth]{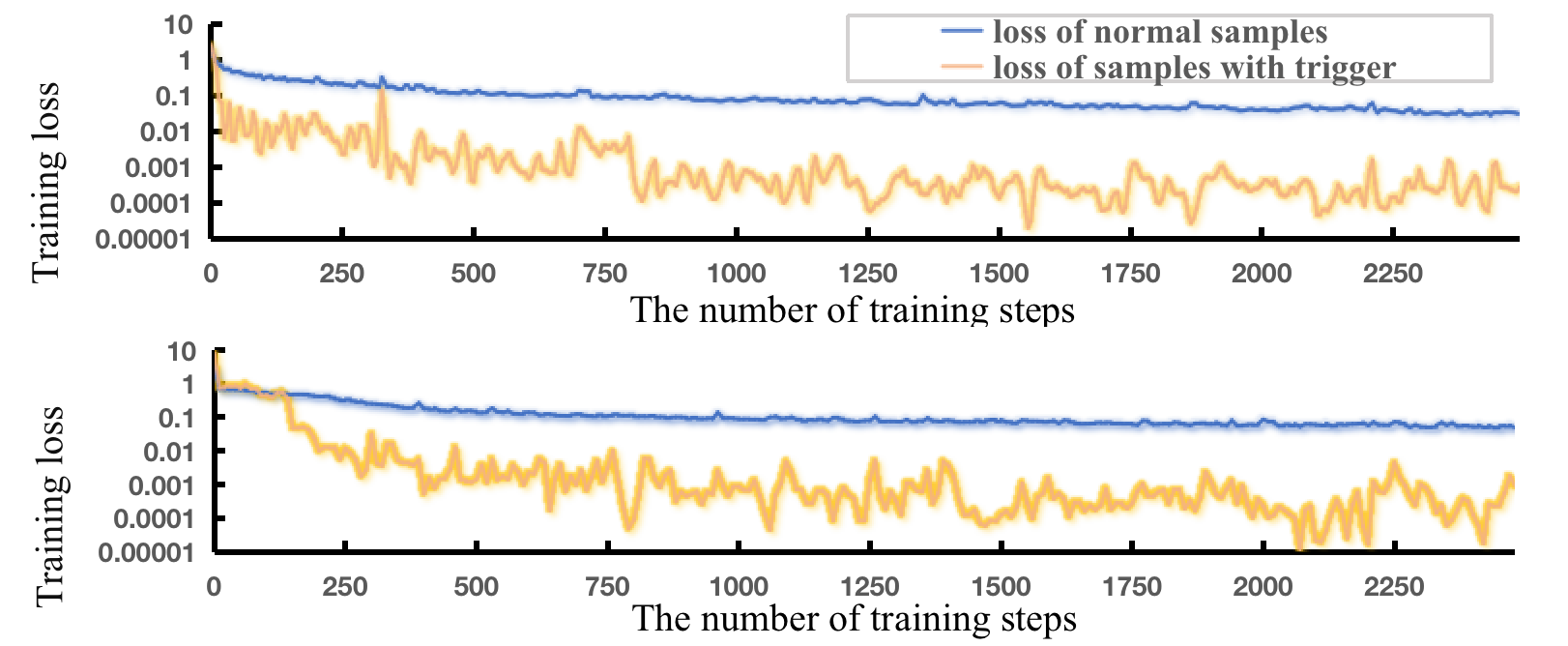}
    \end{center}
       \caption{Training loss for samples w or w/o triggers. The top figure shows the results for MNIST dataset and the bottom figure shows the results for CelebA dataset.}
    \label{data22}
\end{figure}

Afterwards, in both tasks we added 10\% of the trigger to the 
class and recorded the training loss of samples with trigger 
and the normal samples, as shown in Fig.~\ref{data22}.
As can be seen from Fig.~\ref{data22}, the sample loss with a 
trigger quickly drops orders of magnitude, which is different from 
that of the normal training samples. Therefore, at a certain 
stage of training, samples with the trigger make minor 
contribution to the overall gradient of the model, which is 
similar to the effect of deletion.

\subsection{Preliminary Explanation of CAD in Backdoor Attack using Classification Boundaries} 
A study showed that samples with triggers and normal samples 
have a large distance in the feature space and can be easily 
divided into two clusters by a clustering 
algorithm~\cite{DBLP:conf/aaai/ChenCBLELMS19}. 
Based on the experiments in Sec.~\ref{sec21}, above 
study can explain why adding triggers has the same effect 
as deleting them, but it doesn't explain why there are 
differences at higher p\%.
Therefore, we assume that these attacked samples can still 
have a certain impact on the classification boundaries in the 
normal sample cluster in the feature space. 

If features of the backdoor trigger can be easily expressed under 
this assumption, the model will easily determine the class labeled 
with the trigger, so that the backdoor attack retains a high 
Attack Success Rate (ASR). 
Meanwhile, adding triggers to images means moving image samples 
from cluster of normal samples to cluster with triggers in 
feature space. This effect is equivalent to deleting those 
samples from normal samples, and we call the deletion effect 
brought about by backdoor attack as pseudo-deletion.

\begin{figure}[!t]
    \begin{center}
       \includegraphics[width=1.0\linewidth]{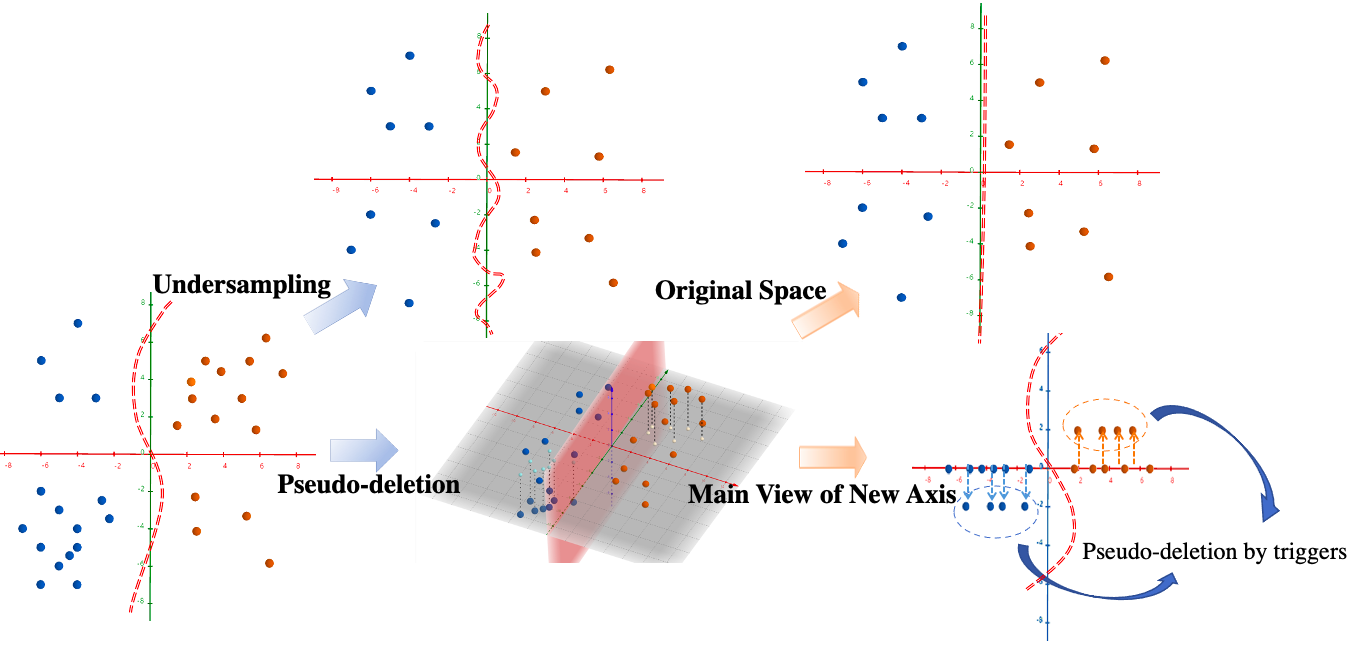}
    \end{center}
    \vspace{-4mm}
    \caption{Schematic diagram of the effect of deletion and
    pseudo-deletion on classification boundaries.}
    \label{data23}
\end{figure}

Although the samples with triggers are far from the cluster 
of the normal samples in the feature space, they still 
satisfy the classification boundary of the classification 
task in the feature space. These samples will also affect 
model's classification boundaries in the normal samples cluster. 
Therefore the experiments in Sec.~\ref{sec21} produced 
generalization differences at higher p\%.
We define the area of the normal samples in 
the feature space as the original space, and the direction 
of moving from the original space to the space with trigger 
samples as the trigger dimension.
Fig.~\ref{data23} shows the original space and 
the trigger dimension.

To verify the above assumption, we designed a toy dataset to 
describe the original feature space. We divided the square 
region into four blocks with 1,000 red and 1,000 blue points 
on the left and right, and set the $bias\ rate$ as the ratio of 
points within the upper left and lower right corners. Points 
in each region are generated from a uniform distribution with 
a small perturbation of the normal distribution. Subsequently, 
we trained a small neural network based on the toy dataset. 
We selected 10,000 points of meshgrid in the sample space to plot the 
classification boundaries, and counted the misclassification 
of those points.

\begin{figure}[!t]
    \begin{center}
        \includegraphics[width=0.7\linewidth]{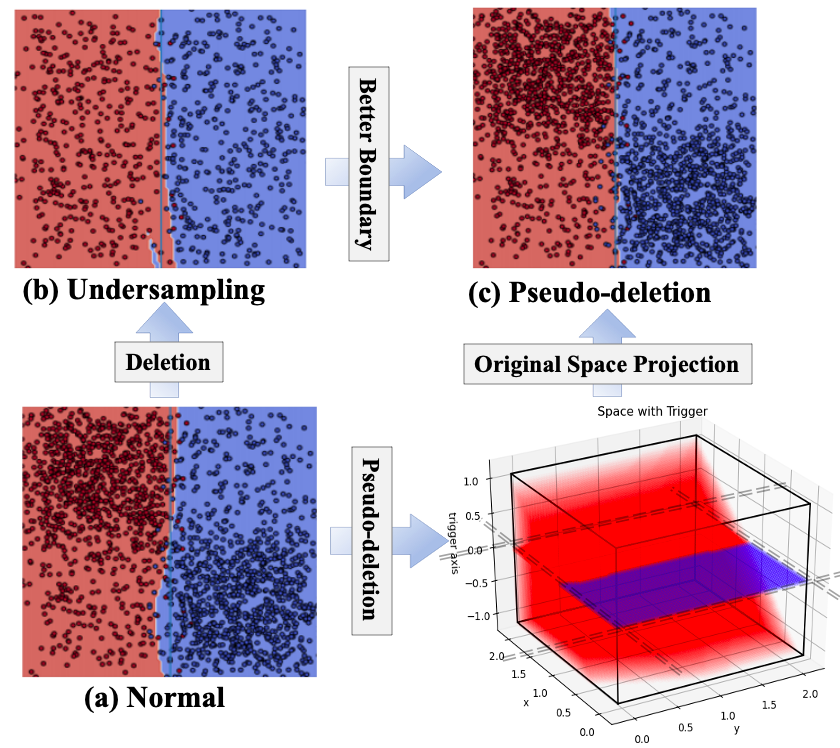}
    \end{center}
    \vspace{-4mm}
    \caption{Visualization of classification boundaries under normal
       training, undersampling, and pseudo-deletion methods when $bias\ rate$ is 0.8.}
    \label{data24}
\end{figure}

\begin{figure}[!t]
    \begin{center}
       \includegraphics[width=0.9\linewidth]{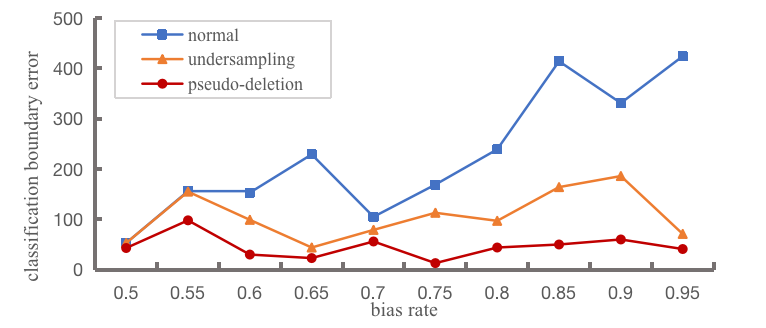}
    \vspace{-4mm}
    \end{center}
       \caption{The value of Classification Boundary Error for normal
       training, undersampling, and pseudo-deletion methods with different
       $bias\ rates$. x-axis is the different $bias\ rates$, and y-axis is the value
       of Classification Boundary Error.}
    \label{data25}
\end{figure}

It can be seen from Fig.~\ref{data24} 
that the model trained on all data causes a significant 
distortion of the classification boundary. Then, we used 
the undersampling method to delete points until the number 
on both sides is balanced. It demonstrates that the accuracy 
of classification boundary is significantly improved but it 
is still not smooth. We also edited the same data by adding 
triggers in a new dimension, and the results hint that this 
method brings a smooth classification boundary. Then, we defined the number of error points in the 
meshgrid as the Classification Boundary Error.
As shown in Fig.~\ref{data25}, we calculated Classification Boundary Error under different $bias\ rates$ and it 
shows that the pseudo-deletion method has the lowest error rate.

To further explore the difference between deletion and 
pseudo-deletion, we compared four methods, namely, 
``delete redundant red and blue points", ``pseudo-delete 
redundant red and blue points", ``delete redundant red 
points and pseudo-delete redundant blue points", and 
``delete redundant blue points and pseudo-delete redundant 
red points".

\begin{figure}[!t]
    \subfigure[]{
        \includegraphics[width=0.22\linewidth]{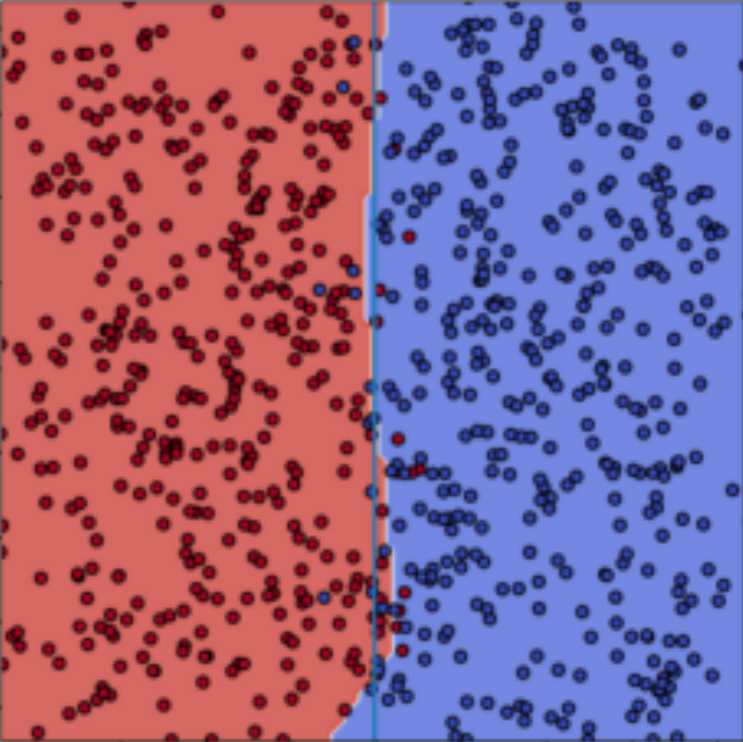}
    }
    \subfigure[]{
        \includegraphics[width=0.22\linewidth]{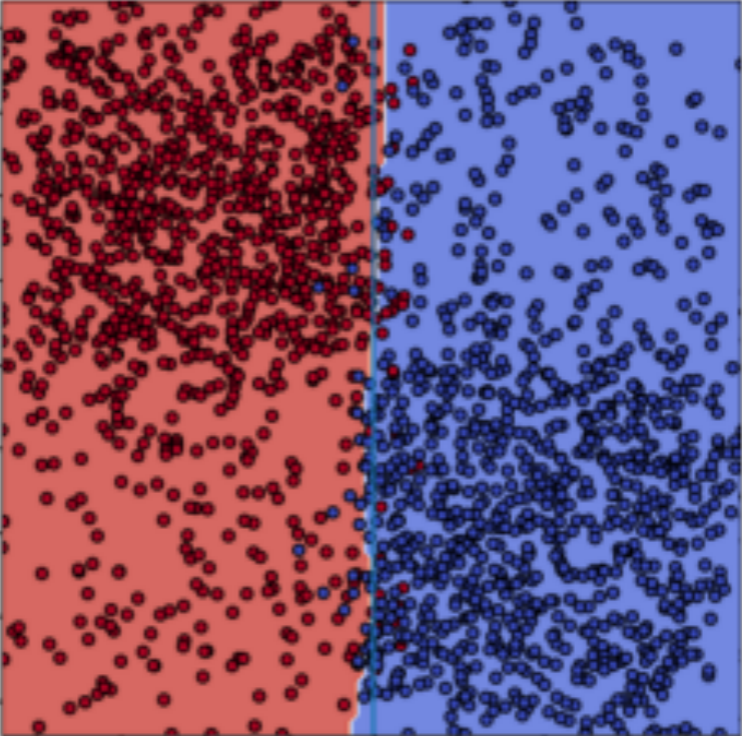}
    }
    \subfigure[]{
        \includegraphics[width=0.22\linewidth]{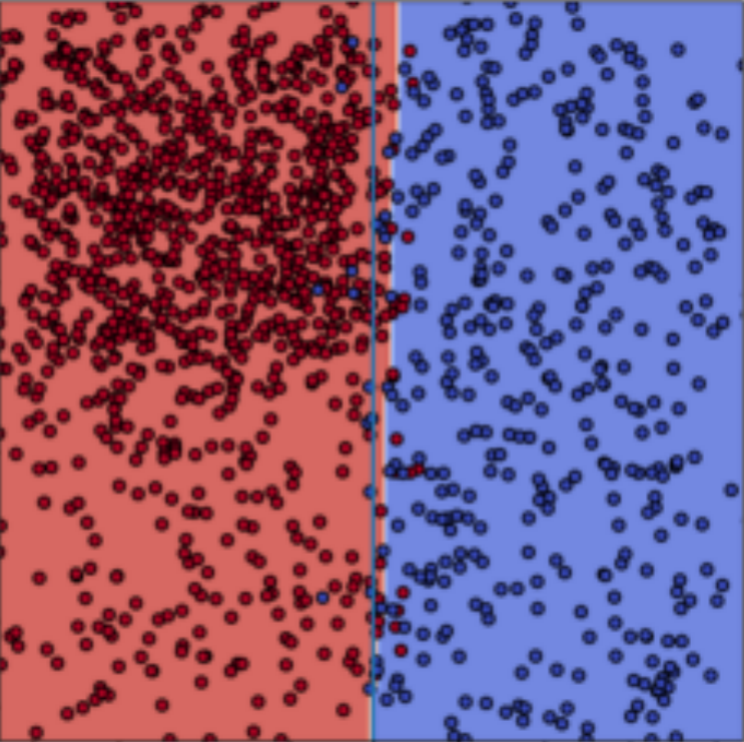}
    }
    \subfigure[]{
        \includegraphics[width=0.22\linewidth]{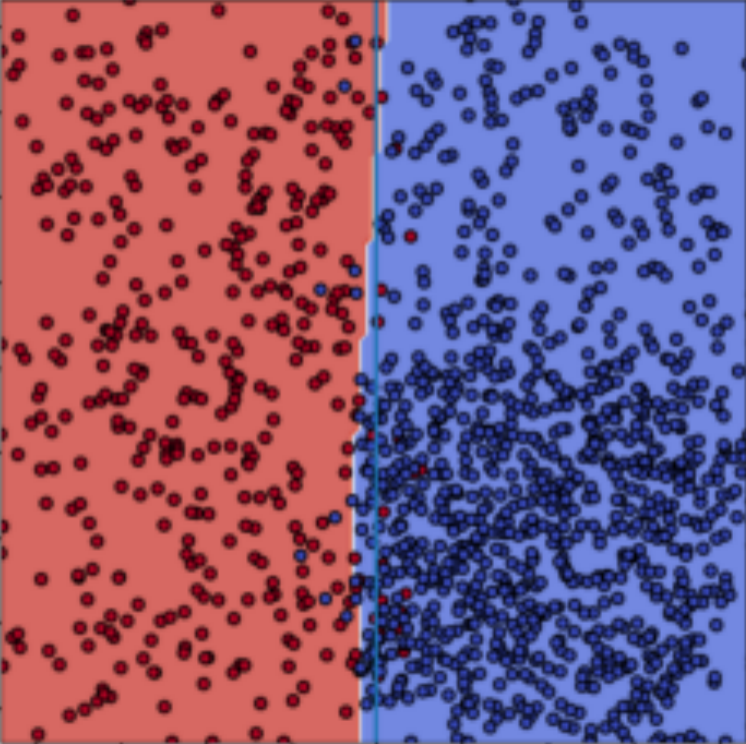}
    }
    \centering
    \resizebox{0.5\linewidth}{!}{
        \begin{tabular}{c|ccc}
            \hline
                   & \multicolumn{3}{c}{Classification Boundary Error} \\ \hline
            Method & left      & right      & total     \\ \hline
            (a)    & 52        & 52         & 104       \\ \hline
            (b)    & 53        & 25         & 78        \\ \hline
            (c)    & 0         & 171        & 171       \\ \hline
            (d)    & 159       & 7          & 166       \\ \hline
            \end{tabular}
    }
       \caption{Visualizing the classification bounds for models trained using all true deletion, all pseudo-deletion, and a mixture of true deletion and pseudo-deletion methods. From left to right: all true deletion, all pseudo-deletion, red points pseudo-deletion blue points true deletion, red points pseudo-deletion blue points true deletion.}
    \label{data26}
\end{figure}

From Fig.~\ref{data26}, the comparison 
between (a) and (b) shows that the classification boundary 
after pseudo-deletion is smoother and closer to the real 
classification boundary. Comparing (c), (d) with (a), we can 
see that the classification boundary in (c) and (d) tends 
to extend towards the real deletion side. Together, the 
present findings confirm that pseudo-deleted samples will 
have an effect on the classification boundary in high 
dimensional space.

\subsection{Debiasing Effect of Backdoor Attack Under Extreme Conditions}
From previous results, we can see that the pseudo-deletion 
based on the backdoor attack is significantly helpful in 
improving classification boundaries. 
Because adding trigger has the effect of deletion, it is 
natural to think that backdoor attack can be combined with 
undersampling methods. Thus, our next goal is to 
test whether it can be applied to model debiasing. 

Undersampling method is an effective and simple method 
in the field of debiasing. Instead of deletion in undersampling, 
pseudo-deletion is considered to test whether the algorithm can 
benefit from it. Our target was to start experiment on CelebA 
dataset. Two variables with the strongest Pearson coefficients 
with gender, $Wearing\_Lipstick$ (Pearson=-0.8196) and 
$Heavy\_Makeup$ (Pearson=-0.6502), were chosen as the 
classification targets with gender bias.

The evaluation is based on Equal Opportunity (Opp.), Equalized 
Odds (Odds) and Equalized Accuracy (EAcc.), which are commonly 
used in the field of debiasing. The above concepts can be 
expressed by True Positive Rate (TPR) and True Negative Rate 
(TNR) as follows:

\begin{equation}
    \begin{small}
        \begin{aligned}
            Opp. &= |TPR_{A=0} - TPR_{A=1}| \\
            Odds &= \frac{1}{2}[|TPR_{A=0} - TPR_{A=1}| + |TNR_{A=0} - TNR_{A=1}|] \\
            EAcc. &= \frac{1}{4}[TPR_{A=0} + TNR_{A=0} + TPR_{A=1} + TNR_{A=1}]
        \end{aligned}
    \end{small}
\end{equation}

\begin{table}[t]
    \resizebox{0.5\textwidth}{!}{
    \begin{tabular}{ccccccc}
        \hline
        \multirow{2}{*}{Method}          & \multirow{2}{*}{Target} & Bais        & \multirow{2}{*}{Opp.$\downarrow$} & \multirow{2}{*}{Odds$\downarrow$} & \multirow{2}{*}{EAcc.$\uparrow$} & \multirow{2}{*}{Acc.$\uparrow$} \\ \cline{3-3}
                                         &                         & M=1 M=0     &                        &                        &                        &                        \\ \hline
        \multirow{2}{*}{Undersampling}   & T=1                     & 51.06 83.73 & \multirow{2}{*}{32.67} & \multirow{2}{*}{22.11} & \multirow{2}{*}{78.46} & \multirow{2}{*}{\textbf{88.10}} \\
                                         & T=0                     & 95.29 83.74 &                        &                        &                        &                        \\ \hline
        \multirow{2}{*}{Pseudo-deletion} & T=1                     & 53.19 78.82 & \multirow{2}{*}{\textbf{25.62}} & \multirow{2}{*}{\textbf{18.09}} & \multirow{2}{*}{\textbf{79.14}} & \multirow{2}{*}{86.72} \\
                                         & T=0                     & 97.55 86.99 &                        &                        &                        &                        \\ \hline
    \end{tabular}
    }
    \caption{The debiasing effect on $Wearing\_Lipstick$ when $Male$ was set as the bias variable.}
    \label{WL}

    \resizebox{0.5\textwidth}{!}{
    \begin{tabular}{ccccccc}
        \hline
        \multirow{2}{*}{Method}          & \multirow{2}{*}{Target} & Bais        & \multirow{2}{*}{Opp.$\downarrow$} & \multirow{2}{*}{Odds$\downarrow$} & \multirow{2}{*}{EAcc.$\uparrow$} & \multirow{2}{*}{Acc.$\uparrow$} \\ \cline{3-3}
                                         &                         & M=1 M=0     &                        &                        &                        &                        \\ \hline
        \multirow{2}{*}{Undersampling}   & T=1                     & 22.73 89.02 & \multirow{2}{*}{66.30} & \multirow{2}{*}{45.89} & \multirow{2}{*}{71.25} & \multirow{2}{*}{89.77} \\
                                         & T=0                     & 99.38 73.88 &                        &                        &                        &                        \\ \hline
        \multirow{2}{*}{Pseudo-deletion} & T=1                     & 40.91 86.38 & \multirow{2}{*}{\textbf{45.47}} & \multirow{2}{*}{\textbf{32.41}} & \multirow{2}{*}{\textbf{76.61}} & \multirow{2}{*}{\textbf{89.93}} \\
                                         & T=0                     & 99.25 79.90 &                        &                        &                        &                        \\ \hline
    \end{tabular}
    }
    \caption{The debiasing effect on $Heavy\_Makeup$ when $Male$ was set as the bias variable.}
    \label{HM}
\end{table}

\begin{figure*}[!t]
    \begin{center}
       \includegraphics[width=1.0\linewidth]{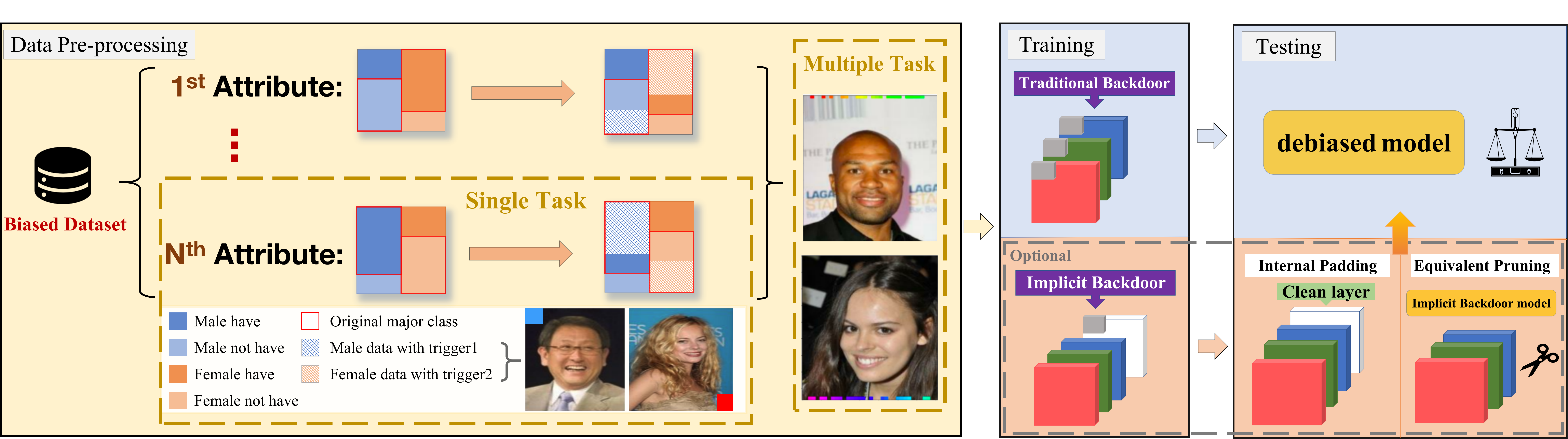}
    \end{center}
       \caption{The proposed framework (using gender bias as example)}
    \label{method31}
\end{figure*}

As shown in Table~\ref{WL} and Table~\ref{HM}, the pseudo-deletion 
method has great potential in debiasing. Compared with undersampling, 
the existing of those pseudo-deleted samples in the feature 
space will make the model to have better performance. In addition, 
pseudo-deletion can keep samples balanced in the dataset same as 
undersampling, and makes the boundary fair and smooth so as to 
achieve debiasing.  

\section{Method}
The undersampling method is often used for data-debiasing by 
deleting large quantities of unbalanced data, sometimes leading 
to non-convergence. By applying pseudo-deletion to undersampling, 
we proposed Debiasing Backdoor Attack (DBA) method 
so as to solve the problem in binary classification with 
single bias variable, and branched out into simultaneously 
classifying multiple binary classification. Besides, 
a framework was brought out in order to describe our method 
while considering to use backdoor attack safely.

\subsection{Debiasing Backdoor Attack for Binary Classification Task}
For binary classification, we replaced data deleting 
with pseudo-deletion in the undersampling method. 
The process can be divided into three steps. 
First, the data distribution in training set was counted, 
based on which the proportion of pseudo-deletion triggers added 
to each sample was calculated as follows:  

\begin{sequation}
    \begin{aligned}
        R(A=i, Y=j) 
         &= \frac{N_{trigger}(A=i, Y=j)}{N(A=i, Y=j)}
        \\ &=\frac{\max[N(A=i, Y=j)-N(A=i, Y=\overline{j}), 0]}{N(A=i, Y=j)} 
    \end{aligned}
\end{sequation}

where $A$ refers to the bias attribute, $Y$ refers to the target 
attribute, $R(\cdot)$ refers to the proportion of the trigger 
added in that case, $N(\cdot)$ refers to the number of training 
samples that meet the case, and similarly $N_{trigger}(\cdot)$ 
refers to the number of samples that need to be added to the 
trigger under that case. Second, add different triggers to 
samples of different bias variables according to the 
calculated proportions, thus generating a 
new training set $\mathcal{D}_{trigger}$. Third, the model was 
trained on the newly generated dataset $\mathcal{D}_{trigger}$ 
to obtain a fair model.

\subsection{For Multi-class Classifications Task}
In this section, we expanded Debiasing Backdoor Attack  
to multi-classification tasks with a single bias variable. 
Multiple binary classification 
tasks can be viewed as N single classification 
tasks. For each single classification task, we 
designed two dedicated triggers, thus there will be 
2N kinds of triggers in $\mathcal{D}_{trigger}$. 
Unlike deleting data, pseudo-deletion triggers 
can be added in the image at the same time as 
much as needed to form trigger barcodes. Therefore, the 
pseudo-deletion method can be used to debias 
multiple binary classification problems, 
as is shown in Fig.~\ref{method31}.

The original undersampling 
for multiple binary classification has to 
take the intersection of each class to make every 
class distribution balanced, and delete the 
rest of large amount of data. Under that condition, 
few data will be last to satisfy the model training. 
In contrast to undersampling methods, Debiasing Backdoor Attack can balance 
distribution easily when applied in the 
multiple binary classification task.

\subsection{Avoiding Security Risks}

In order to avoid security risks of 
backdoor attack, we proposed Implicit Backdoor Attack for training 
and two solutions for testing, namely, Internal Padding 
and Equivalent Pruning.

\noindent\textbf{Implicit Backdoor Attack}: Change RGB channels 
of the image to RGBT channels, by adding a new 
channel T, which is zero by default. The channel T 
is specifically used to adding triggers and other channels 
stay unchanged.

\noindent\textbf{Internal Padding}: During the 
test, users are only allowed to pass in RGB 
three-channel images. The algorithm has to add a T channel on images 
internally with all zeros in order to feed into models, through which 
users cannot edit the T channel so as to keep models safe.

\noindent\textbf{Equivalent Pruning}: Parameters of the first 
layer are edited using the pruning algorithm to 
remove parameters for the T layer so that 
the model only supports RGB input, 
thus avoiding security risks.

\section{Experiments}

\subsection{Experiments Setting}

\subsubsection{Dataset}
\noindent\textbf{CelebA}:
CelebA dataset~\cite{liu2015faceattributes} has 200k face images and is annotated with 40 attributes. 
Male~\footnote{Since the attributes $5\_o\_Clock\_Shadow$, $Goatee$, $Mustache$, $No\_Beard$, and $Sideburns$ are not suitable for studying gender bias, we do not discuss these attributes in our subsequent experiments.} 
and Young attributes were set as bias variables and the rest as target attributes.
The training set split method in tensorflow-dataset was applied.

\noindent\textbf{UTK Face}:
UTK Face dataset~\cite{zhifei2017cvpr} has 23,705 face images and is labeled with 3 attributes (age, gender, and race). We set the ratio of training set to test set as 7:3, and chose $Race$ (whether or not white race) as the bias attribute to predict whether people were older than 35.

\subsubsection{Detail Setting}
For all algorithms we set ResNet9 as the backbone network, selected Adam as the optimizer, and set the learning rate to 0.001. For multiple binary classification prediction tasks, we set ResNet9 to have multiple linear output heads, with each head used to predict one attribute.
We chose Opp., Odds, EAcc., and Acc. to evaluate algorithms, and do not record the value of Opp. and Odds if the algorithm does not converge.

\subsubsection{Comparing Models}
Typical debiasing alternatives are supplied for comparison:
\begin{normalsize}
\begin{itemize}
    \item Undersampling~\cite{DBLP:journals/tkde/ZhouL06}: discarding samples with majority bias variable to construct a balanced dataset.
    \item Reweighting~\cite{DBLP:journals/kais/KamiranC11}: assigning different weights to samples and modifying the training objectives to softly balance the data distribution.
    \item Adversarial Learning~\cite{DBLP:conf/icml/GaninL15,wadsworth2018achieving}: the typical in-processing debiasing solution by adversarially learning between the target and bias tasks.
    \item Fairness GAN~\cite{DBLP:journals/ibmrd/SattigeriHCV19}: an auxiliary classifier GAN that strives for equality of opportunity.
    \item FFVAE~\cite{DBLP:conf/icml/CreagerMJWSPZ19}: learning compact representations that are useful for reconstruction and fair prediction.
    \item TAC~\cite{DBLP:conf/accv/HwangPLJ0B20}: the method attempts to train the fairness-aware image classification model without protected attribute annotations.
    \item MFD~\cite{Jung_2021_CVPR}: a systematic approach which reduces algorithmic biases via feature distillation for visual recognition tasks.
    \item FD-VAE~\cite{DBLP:conf/aaai/ParkH0B21}: a VAE algorithm that learns fair representations by decomposing the data representation into three independent subspaces.
    \item Fair Mixup~\cite{DBLP:conf/iclr/ChuangM21}: a data augmentation strategy for imposing the fairness constraint.
    \item BiFair~\cite{DBLP:journals/corr/abs-2106-04757}: a training algorithm that can jointly minimize for a utility and a fairness loss.
    \item RNF-GT~\cite{DBLP:journals/corr/abs-2106-12674}: discouraging the classification head from capturing spurious correlation between fairness sensitive information in encoder representations with specific class labels.
\end{itemize}
\end{normalsize}

\subsection{Debiasing Effect}

\begin{table}[t]
    \resizebox{0.5\textwidth}{!}{
    \begin{tabular}{cllll}
    \hline
    Method                & Opp.$\downarrow$         & Odds$\downarrow$         & EAcc.$\uparrow$          & Acc.$\uparrow$          \\ \hline
    Normal                & 21.28         & 15.21         & 74.99          & \textbf{87.17} \\ \hline
    Undersampling            & 6.96          & 6.99          & 79.86          & 79.22          \\ \hline
    Reweighting           & 10.00         & 7.81          & 75.47          & 78.83          \\ \hline
    Adversarial Learning & 17.38         & 14.02         & 71.82          & 85.98          \\ \hline
    DBA(Ours)                  & \textbf{4.73} & \textbf{5.58} & \textbf{81.21} & 82.13          \\ \hline
    \end{tabular}
    } 
    \caption{The average effect of different algorithms for debiasing the remaining 34 attributes in CelebA dataset with $Male$ as the bias variable.}
    \label{male}
\end{table}

\begin{table}[t]
    \resizebox{0.5\textwidth}{!}{
    \begin{tabular}{cllll}
    \hline
    Method                & Opp.$\downarrow$         & Odds$\downarrow$         & EAcc.$\uparrow$          & Acc.$\uparrow$          \\ \hline
    Normal                & 7.27          & 6.05          & 80.22          & \textbf{89.77} \\ \hline
    Undersampling            & 3.54          & 5.09          & 82.95          & 83.67          \\ \hline
    Reweighting           & 5.13          & 5.01          & 78.10          & 81.67          \\ \hline
    Adversarial Learning & 6.45          & 5.62          & 79.01          & 88.12          \\ \hline
    DBA(Ours)                  & \textbf{1.63} & \textbf{4.30} & \textbf{83.99} & 82.12          \\ \hline
    \end{tabular}
    }
    \caption{The average effect of different algorithms for debiasing the remaining 39 attributes in CelebA dataset with $Young$ as the bias variable.}
    \label{young}
\end{table}

\begin{table}[t]
    \resizebox{0.5\textwidth}{!}{
        \begin{tabular}{ccccccc}
            \hline
            \multirow{2}{*}{Method}               & \multirow{2}{*}{Target} & Bais        & \multirow{2}{*}{Opp.$\downarrow$} & \multirow{2}{*}{Odds$\downarrow$} & \multirow{2}{*}{EAcc.$\uparrow$} & \multirow{2}{*}{Acc.$\uparrow$} \\ \cline{3-3}
                                                  &                         & R=1 R=0     &                        &                        &                        &                        \\ \hline
            \multirow{2}{*}{Normal}               & T=1                     & 74.17 57.16 & \multirow{2}{*}{17.01} & \multirow{2}{*}{11.54} & \multirow{2}{*}{77.44} & \multirow{2}{*}{\textbf{81.60}} \\
                                                  & T=0                     & 86.18 92.25 &                        &                        &                        &                        \\ \hline
            \multirow{2}{*}{Undersampling}           & T=1                     & 74.82 68.35 & \multirow{2}{*}{6.47}  & \multirow{2}{*}{4.31}  & \multirow{2}{*}{76.00} & \multirow{2}{*}{77.60} \\
                                                  & T=0                     & 79.35 81.50 &                        &                        &                        &                        \\ \hline
            \multirow{2}{*}{Reweighting}          & T=1                     & 0.00 0.00   & \multirow{2}{*}{——}  & \multirow{2}{*}{——}  & \multirow{2}{*}{50.00} & \multirow{2}{*}{63.62} \\
                                                  & T=0                     & 100.00 100.00 &                      &                        &                        &                        \\ \hline
            \multirow{2}{*}{Adversarial Learning} & T=1                     & 93.71 81.91 & \multirow{2}{*}{11.80} & \multirow{2}{*}{10.92} & \multirow{2}{*}{79.51} & \multirow{2}{*}{78.47} \\
                                                  & T=0                     & 66.19 76.23 &                        &                        &                        &                        \\ \hline
            \multirow{2}{*}{DBA(Ours)}     & T=1                     & 83.07 78.36 & \multirow{2}{*}{\textbf{4.71}}  & \multirow{2}{*}{\textbf{3.50}}  & \multirow{2}{*}{\textbf{80.32}} & \multirow{2}{*}{80.07} \\
                                                  & T=0                     & 81.07 78.78 &                        &                        &                        &                        \\ \hline

        \end{tabular}
        }
        \caption{The effect of debiasing different algorithms on whether the age is older than 35 in UTK Face dataset with $Race$ as the bias variable.}
        \label{utk}
\end{table}

First, we conducted experiments on the classification of single 
attributes and reproduced four commonly used algorithms, 
namely, Normal, Undersampling, Reweighting and Adversarial Learning. 
In Table~\ref{male}, $Male$ was set as the bias variable. 
Opp., Odds, EAcc. and Acc. are the average of remaining 34 groups 
attributes in CelebA. It shows that our proposed 
method has a huge debiasing advantage over the above mentioned 
four methods. Although a decline in the accuracy, 
Opp., Odds, EAcc. are significantly improved. In Table~\ref{young},
$Young$ was set as another bias variable
for the remaining 39 sets of attribute. 
It shows that the score of our algorithm on Opp.
is half lower than that of Undersampling on age bias. To avoid the 
error caused by the dataset, $Race$ was set as the bias variable 
on UTK Face to predict whether the age was older than 35. 
As shown in Table~\ref{utk}, under the extreme 
bias of $Race$, our algorithm still has good results 
while the Reweighting algorithm appears to be non-convergent.

\begin{table}[t]
    \centering
    \resizebox{0.45\linewidth}{!}{
        \begin{tabular}{cc}
        \hline
        Method               & Num \\ \hline
        Normal               & \textbf{0}   \\ \hline
        Undersampling          & 2   \\ \hline
        Reweighting          & 12  \\ \hline
        Adversarial Learning & 3   \\ \hline
        DBA(Ours)     & \textbf{0}   \\ \hline
        \end{tabular}
    }
    \caption{Counts of non-convergence cases for different algorithms in the above 74 sets of experiments.}
    \label{convergence}
\end{table}

The frequencies of non-convergence 
of each algorithm in the 74 sets of experiments are shown in 
Table~\ref{convergence}. In extreme cases, 
the Reweight algorithm is most prone 
to non-convergence due to loss imbalance. The Undersampling 
algorithm will also fail to converge due to the deletion of 
excess data. 
The Adversarial Learning will not converge for its 
complex training process.
Compared with those methods, our method has strong applicability under 
different conditions.

Second, we looked for eight new debiasing methods with 
the same experimental settings as ours and extracted their 
best results for a fair comparison. As is shown in  
Table~\ref{8method}, our method has so far achieved SOTA in debiasing.

\begin{table}[t]
    \resizebox{1.0\linewidth}{!}{
    \begin{tabular}{cccccc}
    \hline
    Method                           & Bias-Target     & Opp.$\downarrow$         & Odds$\downarrow$         & EAcc.$\uparrow$          & Acc.$\uparrow$   \\ \hline
    \hline
    Fairness GAN                     & Male-Attractive & 22.53 & 5.07  & 73.43 & -     \\ \hline
    FFVAE                            & Male-Attractive & 15.00 & 16.60 & 59.20 & 62.20 \\ \hline
    TAC                              & Male-Attractive & 10.23 & -     & -     & -     \\ \hline
    MFD                              & Male-Attractive & -     & 5.46  & -     & \textbf{80.15} \\ \hline
    FD-VAE                           & Male-Attractive & 1.30  & 4.90  & 63.50 & 64.10 \\ \hline
    Fair Mixup                       & Male-Attractive & -     & 4.50  & -     & -     \\ \hline
    DBA(Ours)                 & Male-Attractive & \textbf{0.65}  & \textbf{4.00}  & \textbf{77.95} & 78.08 \\ \hline
    \hline
    BiFair                           & Male-Smile      & -     & 2.50  & -     & -     \\ \hline
    DBA(Ours)                 & Male-Smile      & \textbf{1.33}  & \textbf{2.06}  & \textbf{91.01} & \textbf{91.18} \\ \hline
    \hline
    RNF-GT                           & Male-Wavy\_Hair & -     & 15.00 & -     & 70.80 \\ \hline
    DBA(Ours)                 & Male-Wavy\_Hair & \textbf{9.19}  & \textbf{8.10}  & \textbf{73.58} & \textbf{72.77} \\ \hline
    \end{tabular}
    }
    \caption{
        Comparison results between our method and 8 methods in the same experimental configuration.
    }
    \label{8method}
    \end{table}

\begin{table}[t]
    \centering
    \resizebox{0.75\linewidth}{!}{
    \begin{tabular}{ccccc}
    \hline
    Method                & Opp.$\downarrow$         & Odds$\downarrow$         & EAcc.$\uparrow$          & Acc.$\uparrow$          \\ \hline
    Normal(Male)                & 22.54         & 15.38         & 70.44          & \textbf{86.74} \\ \hline
    DBA(Male)                  & \textbf{12.73} & \textbf{10.83} & \textbf{73.43} & 82.97         \\ \hline
    \end{tabular}
    }
    \centering
    \resizebox{0.75\linewidth}{!}{
    \begin{tabular}{cllll}
    \hline
    Method                & Opp.$\downarrow$         & Odds$\downarrow$         & EAcc.$\uparrow$          & Acc.$\uparrow$          \\ \hline
    Normal(Young)                & 6.64          & 5.11          & 74.08          & \textbf{88.77} \\ \hline
    DBA(Young)                  & \textbf{4.83} & \textbf{6.14} & \textbf{80.44} & 84.44          \\ \hline
    \end{tabular}
    }
    \caption{The average effect of a special debiased scene for classifying multiple attributes at the same time.}
    \label{muti}
\end{table}

Specially, our algorithm can be applied in debiasing 
multiple attributes simultaneously. 
In Table~\ref{muti}, our algorithm for classifying multiple 
attributes performs better than average results of 
Adversarial Learning and Reweighting in 
classifying a single attribute.

\subsection{Ablation Experiments}

Ablation experiments aim to test whether different 
patch sizes and colors will successfully achieve 
pseudo-deletion. Results are shown in Table~\ref{tab4}.

Unlike backdoor attack methods that 
consider invisibility, the color and size of the trigger will 
not affect the results of the backdoor attack when a simple 
style of trigger is used.

\begin{table}[!t]
    \centering
    \resizebox{0.55\linewidth}{!}{
    \begin{tabular}{cc}    
        \hline
        Experiments Setting      & Results \\ \hline
        Red Patch, 25pix*25pix   & Success \\ \hline
        Blue Patch, 25pix*25pix  & Success \\ \hline
        Red Patch, 10pix*10pix   & Success \\ \hline
        Red Patch, 5pix*5pix     & Success \\ \hline
    \end{tabular}
    }
    \caption{Results of whether patch works with different experimental settings.}
    \label{tab4}
\end{table}

\subsection{Results of Avoiding Security Risks}
Results of EAcc., parameter 
numbers and resources costing 
among normal Debiasing Backdoor Attack, Internal Padding and 
Equivalence Pruning are shown in Table~\ref{tab5}.

\begin{table}[!t]
    \centering
    \resizebox{1.0\linewidth}{!}{
    \begin{tabular}{cccc}    
        \hline
        Method & EAcc. & Params & Cost \\ \hline
        Normal & 81.21 & 4.908M & 1.472 GFlops\\ \hline
        Implicit + Internal Padding & 81.21 & 4.911M & 1.533 GFlops\\ \hline
        Implicit + Equivalence Pruning & 81.21 & 4.908M & 1.472 GFlops\\ \hline
    \end{tabular}
    }
    \caption{Comparison of different methods for avoiding security risks in terms of EAcc., number of parameters, and computational resource usage.}
    \label{tab5}
\end{table}

As for as EAcc., there is no difference 
between the three methods, whereas the normal method has 
security risks. Compared with the other two safe methods, 
the Equivalence Pruning perform better 
which is suitable for model deployments. 
However, it is difficult to handle.
Meanwhile, Internal Padding fits for 
testing in experimental environments.

\section{Conclusions}
In summary, starting from the observation on CAD, 
we found that the backdoor attack, as a substitute for 
data deletion methods, had the potential to eliminate 
data bias. On top of that, we believe 
other phenomena in backdoor attack are worth discussion, 
for instance, by studying the learning speed of trigger, 
we may obtain some properties such as 
the learning order of model features. 
As a way of model probe to understanding model properties, 
backdoor attack is of considerable value to exploring the 
mechanism of deep learning and offering alternative 
solutions to notorious problems.

\appendix

\bibliographystyle{named}
\bibliography{ijcai22}

\end{document}